# Modeling Bike Counts in a Bike-Sharing System Considering the Effect of Weather Conditions

Huthaifa I. Ashqar
Charles E. Via, Jr. Department of Civil and Environmental Engineering
Virginia Tech Transportation Institute
3500 Transportation Research Plaza, Blacksburg, VA 24061
hiashqar@vt.edu

Mohammed Elhenawy, Ph.D.
Virginia Tech Transportation Institute
3500 Transportation Research Plaza, Blacksburg, VA 24061
Tel: 540-231-0278
Fax: 540-231-1555
mohame1@vt.edu

Hesham A. Rakha, Ph.D., P.Eng. (Corresponding author)
Charles E. Via, Jr. Department of Civil and Environmental Engineering
Virginia Tech Transportation Institute
3500 Transportation Research Plaza, Blacksburg, VA 24061
Tel: 540-231-1505
Fax: 540-231-1555
hrakha@vt.edu

# Modeling Bike Counts in a Bike-Sharing System Considering the Effect of Weather Conditions

## 1 Abstract

The paper develops a method that quantifies the effect of weather conditions on the prediction of bike station counts in the San Francisco Bay Area Bike Share System. The Random Forest technique was used to rank the predictors that were then used to develop a regression model using a guided forward step-wise regression approach. The Bayesian Information Criterion was used in the development and comparison of the various prediction models. We demonstrated that the proposed approach is promising to quantify the effect of various features on a large BSS and on each station in cases of large networks with big data. The results show that the time-of-the-day, temperature, and humidity level (which has not been studied before) are significant count predictors. It also shows that as weather variables are geographic location dependent and thus should be quantified before using them in modeling. Further, findings show that the number of available bikes at station $i$ at time $t-1$ and time-of-the-day were the most significant variables in estimating the bike counts at station $i$.

**Keywords:** Bike Counts Prediction, Bike-Sharing, Big Data, Random Forest, Urban Computing.

# 2 Introduction

A growing population, with more people living in cities, has led to increased pollution, noise, congestion, and greenhouse gas emissions. One possible approach to mitigating these problems is encouraging the use of bikes through bike sharing systems (BSSs). BSSs are an important part of urban mobility in many cities and are sustainable and environmentally-friendly systems. As urban density and its related problems increase, it is likely that more BSSs will exist in the future. The relatively low capital and operational cost, ease of installation, existence of pedal assistance for people who are physically unable to pedal for long distances or on difficult terrain, and better tracking of bikes are some of the properties that strengthen this prediction [1].

One of the first BSSs in the United States came into existence in 1994 with a small bike sharing program in Portland, which had only 60 bicycles available for public use. At present, although the BSS experience is still relatively limited, many cities, such as San Francisco and New York, have launched programs to serve users using different payment structures and conditions. One of the largest information technology (IT)-based systems, based in Montreal, Canada, is BIXI (BIcycle-TaXI) that uses a bicycle as a taxi. In fact, this system, with its use of advanced technologies for implementation and management, demonstrates a shift into the fourth generation of BSSs [2].

In 2013, San Francisco launched the Bay Area BSS, a membership-based system providing 24 hours a day, 7 days a week self-service access to short-term rental bicycles. Members can check out a bicycle from a network of automated stations, ride to the station nearest their destination, and leave the bicycle safely locked for someone else to use [3]. The Bay Area BSS is designed for short, quick trips, and as a result, additional fees apply to trips longer than 30 minutes. In this system, 70 bike stations connect users to transit, businesses and other destinations in four different major areas: downtown San Francisco, Palo Alto, Mountain View, and downtown San Jose [3]. The Bay Area BSS is available to everyone 18 years and older with a credit or debit card. The system is designed to be used by commuters and tourists alike, whether they are trying to get across town during the rush hour, traveling to and from the Bay Area Rapid Transit (BART) system and Caltrain stations, or for any other daily activities [3].

This paper proposes an approach to constructing a bike count model for the San Francisco Bay Area BSS. The bike counts at each station, each of which has a finite number of docks, fluctuates. Thus, a rebalancing (or redistribution) operation must be performed periodically to meet this fluctuation. Coordinating such a large operation is complicated, time consuming, polluting and expensive [1]. Firstly, this paper attempts to quantify the effect of several variables on the mean of bike counts for the Bay Area BSS network, including the month-of-the-year, the day-of-the-week, time-of-the-day, and various weather conditions. Secondly, using the same proposed method, the paper construct a predictive model for the bike counts at each station over the time as it is one of the key tasks to making the rebalancing operation more efficient.

In terms of the paper layout, following the introduction, this paper is organized into six sections. First, related work, focused on the proposed model in previous studies, is discussed. Next, a background of count model regression, Random Forest, and Bayesian Information Criterion are presented. Third, the different data sets used in this study are described. In the fourth section, the details of the data analysis used to quantify the effect of various features in BSS are discussed. Next, results of constructing a predictive bike count model are provided. Finally, the paper concludes with a summary of new insights and recommendations for future bike count model research.

## 3   Related Work

The modeling of BSS data using various features, including time, weather, built-environment, transportation infrastructure, etc., is an area of significant research interest. In general, the main goals of data modeling are to boost the redistribution operation [4-8], to gain new insights into and correlations between bike demand and other factors [9-15], and to support policy makers and managers in making good decisions [9, 16-18]. Generally, the main approach to modeling and predicting bike sharing data is regression count modeling. A recent paper modeled the demand for bikes and return boxes using data from the BSS Citybike Wien in Vienna, Austria. The influence of weather (temperature and precipitation) and full/empty neighboring stations on demand was studied using different count models (Poisson, Negative Binomial [NB] and Hurdle). The authors found that although the Hurdle model worked best in modeling the demand of bike sharing stations, these models were complex and might not be ideal for optimization procedures. They also found that NB models outperformed Poisson models because of the dispersion in the data (to be discussed later) [11]. However, an early study used count series to predict the stations' usage based on Poisson mixtures, providing insight into the relationship between station neighborhood type and mobility patterns [19].

In a study by Wang et al., log-linear and NB regression models were used to estimate total station activity counts. The factors used included: economical, built-environment, transportation infrastructural and social aspects, such as neighborhood sociodemographic (i.e., age and race), proximity to the central business district, proximity to water, accessibility to trails, distance to other bike share stations, and measures of economic activity. All the variables were found to be significant. The Log-likelihood was used as a measure of the goodness of fit of the Poisson and NB models [10]. Linear least squared regression with data from the on-the-ground Capital BSS was implemented in another paper to predict station demand based on demographic, socioeconomic, and built-environment characteristics [9].

Several studies used methods other than count models to model BSS data. A multivariate linear regression analysis was used in another study to study station-level BSS ridership. That study investigated the correlation between BSS ridership and the following factors: population density; retail job density; bike, walk, and transit commuters; median income; education; presence of bikeways; nonwhite population (negative association); days of precipitation (negative association); and proximity to a network of other BSS stations. The authors found that the demographic, built environment, and access to a comprehensive network of stations were critical factors in supporting ridership [12].

A study by Gallop et al. used continuous and year-round hourly bicycle counts and weather data to model bicycle traffic in Vancouver, Canada. The study used seasonal autoregressive integrated moving average analysis to account for the complex serial correlation patterns in the error terms and tested the model against actual bicycle traffic counts. The study demonstrated that the weather had a significant and important impact on bike usage. The authors found that the weather data (namely temperature, rain, humidity, and clearness) were generally significant; temperature and rain, specifically, had an important effect [20].

It is also worth noting that some studies used methods other than regression to either model BSS data or to develop new insights and understandings of BSSs (see [4, 16]). For example, a mathematical formulation for the dynamic public bike-sharing balancing problem was introduced using two different models: the arc-flow formulation and the Dantzig-Wolfe decomposition formulation. The demand was computed by considering the station either a pickup or delivery point, with a real-time and length period between two stations [4].

## 4  Methods
### 4.1  Count Models

In the model used for this study, the outcomes $y_i$ (bike count in our prediction model) are discrete non-negative integers, and they represent the number of available bikes at a specified time at each station in the network. Count models based on generalized linear models (GLMs) were applied. Specifically, two models were used to predict the bike count in the network: the Poisson regression model (PRM) and the Negative Binomial regression model (NBRM). Following are brief descriptions of these two models; more details can be found in the literature [21, 22].

#### 4.1.1  Poisson Regression Model (PRM)

In the PRM, each observation $i$ is allowed to have a different mean $\mu$, where $\mu_i$ is estimated from recorded characteristics. The PRM assumes that $y$ has a Poisson distribution, and its logarithm (i.e., link function) can be modeled as a linear combination of parameters. However, the Poisson distribution assumes that the mean and variance are equal $Var(y) = \mu$. If this condition is not met, there is an over-dispersion in the data, implying that more complex models need to be applied. The probability density for the PRM is

$$f(y, \mu) = \frac{\exp(-\mu)\, \mu^y}{y!} \qquad (1)$$

The GLM of the mean $\mu$ on a predictor vector $x_i$ is formulated as

$$\log(\mu_i) = \beta_i x_i^T \tag{2}$$

where $\beta_i$ are the estimated regression coefficients and $\log(\mu_i)$ is the natural logarithm.

### 4.1.2 Negative Binomial Regression Model (NBRM)

The NBRM is considered a generalization of PRM. It is based on a Poisson-gamma mixture distribution that assumes that the count $y_i$ is dependent on two parameters: the mean $\mu_i$ and some dispersion parameter $\theta$. It basically loosens the assumption in PRM that the variance is equal to the mean and adjusts the variance independently. In fact, the Poisson distribution is a special case of the Negative Binomial distribution. The probability density for the NBRM is

$$f(y,\mu) = \frac{\Gamma(y+\theta)}{\Gamma(\theta)y!} \frac{\mu^y \theta^\theta}{(\mu+\theta)^{y+\theta}} \tag{3}$$

The GLM of the mean $\mu$ on the predictor vector $x_i$ is formulated as

$$\log(\mu_i) = \beta_i x_i^T \tag{4}$$

where $\beta_i$ are the estimated regression coefficients and $\log(\mu_i)$ is the natural logarithm.

### 4.1.3 PRM vs NBRM

The Poisson distribution assumes that the mean and variance are the same. However, occasionally, the data shows that the variance might be higher or lower than the mean. This situation is called over-dispersion/under-dispersion and NBRM is able to accommodate these cases. The NB distribution has an additional parameter to the Poisson distribution, which adjusts the variance independently from the mean. In fact, the Poisson distribution is a special case of the negative binomial distribution. Thus, the PRM and the NBRM have the same mean structure, but the NBRM has one parameter more than the PRM to regulate the variance independently from the mean. As Cameron and Trivedi explain, *"if the assumptions of the NBRM are correct, the expected rate for a given level of the independent variables will be the same in both models. However, the standard errors in the PRM will be biased downward, resulting in spuriously large z-values and spuriously small p-values"* [21, 23].

## 4.2 Random Forest (RF)

One of the characteristics of this type of data set is that it is often very large. It is therefore crucial to implement machine learning to identify potential explanatory variables [16]. Moreover, when a model contains a large number of predictors it becomes more complex and overfitting can occur. To avoid this, the Random Forest (RF), as introduced by Breiman in 2001 [24], was applied. The RF creates an ensemble of decision trees and randomly selects a subset of features to grow each tree. While the tree is being grown, the data are divided by employing a criterion in several steps or nodes. The correlation between any two trees and the strength of each individual tree in the forest, also known as the forest error rate in classifying each tree, affect the model. Practically, the mean squared error of the responses is used for regression. The RF method randomly constructs a collection of decision trees in which each tree chooses a subset of features to grow and, then, the results are obtained based on the majority votes from all trees. The number of decision trees and the selected features for each tree are user-defined parameters. The reason for choosing only a subset of features for each tree is to prevent the trees from being correlated.

The fact that in the RF each tree is constructed using a different bootstrap sample from the original data ensures that the RF extracts an unbiased estimate of the generalization error. This is called the OOB (out-of-bag) error estimate, which can be used for model selection and validation without the need for a separate test. The OOB was used to validate the significance of the subsequent inference of each parameter in this study. The RF technique offers several advantages. For example, it offers protection against the impact of collinearity between predictors by building bagged tree ensembles and randomly choosing a subset of features for each tree in a random forest; it runs efficiently with a large amount of data and many input variables without the need to create extra dummy variables; it can handle highly nonlinear variables and categorical interactions; and it ranks each variable's individual contributions in the model. However, RF also has a few limitations. For instance, the observations must be independent, which is assumed in our case. Moreover, model interpretation after averaging many tree models is generally more difficult than interpreting a single-tree model. However, this is not relevant to our model, as it was used only for ranking the predictors. For more details see [24-26].

In this study, RF was used as a technique to rank the effect of the different parameters in the model. This rank was used as a systematic guide in the forward step-wise technique. Performing a direct stepwise regression for a BSS is difficult, as there are many predictors involved in the process, which is time consuming, expensive, and requires expensive statistical software (for example, see [9]). Therefore, we employed the Bayesian Information Criterion (BIC) (discussed in the next section) to choose the most accurate model while maintaining model simplicity. We started by modeling the most important parameters using RF as the only explanatory variable (aka the regressor). Then, forward step-wise regression was applied and the log-likelihood was found and applied to determine the accumulated BIC.

### 4.3   Bayesian Information Criterion (BIC)

BIC was the criterion selected to compare between models following a forward step-wise regression guided by the results of RF. In general, the model with the lowest BIC is preferred.

Adding predictors may increase the log-likelihood, leading to overfitting, and log-likelihood does not take into account the number of predictors. BIC makes up for the number of predictors in the model by introducing a penalty term. Given that $\hat{L}$ is the maximum likelihood, $n$ is number of observations, and $k$ is the number of predictors, BIC is defined as [27]

$$BIC = -2.\ln\hat{L} + k.\ln(n) \qquad (5)$$

As shown in Equation (5), $k.\ln(n)$ is the term to account for the number of predictors in each model.

## 5 Data Set

This study used anonymized bike trip data collected from August 2013 to August 2015 in the San Francisco Bay Area as shown in Figure 2 [28]. This study used two data sets. The first data set included the station ID, number of bikes available, number of docks available, and time of recording. The time data included the year, month, day-of-the-month, time-of-the-day, and minutes at which an incident was recorded. As an incident was documented every minute for 70 stations in San Francisco over 2 years, this data set contained a large number of recorded incidents. This data set was exposed to a change detection process to determine times when a change in bike count occurred at each station. From this data set, as a result of pre-processing, the station ID, number of bikes available, month, day-of-the-week, and time-of-the-day were extracted for use as a feature. Time-of-the-day is considered as the time resolution of the bike counts and was regressed as 0:23; i.e. hours in a day. Subsequently, each station's ZIP code was assigned and input to the set. Figure 1 shows a histogram of the bike counts for all stations resulting from the change detection process. The histogram is considerably skewed to the right, which means that the mean, median, and mode are markedly different, indicating a dispersion in the counts.

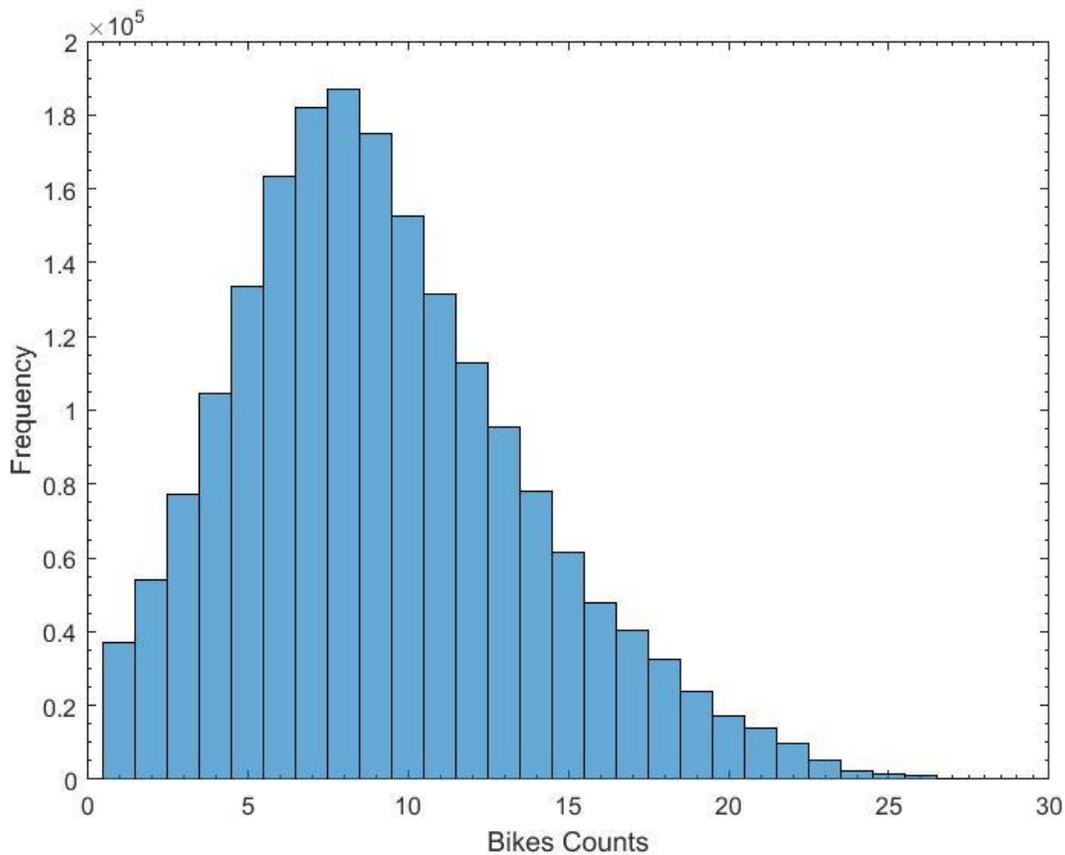

*Figure 1. Histogram of the bike counts*

The second data set contained different attributes: the date (in month/day/year format), ZIP code, and other variables describing the daily weather for each ZIP code over the 2-year period. Daily weather data at each ZIP code contained information about the temperature, humidity, dew level, sea level pressure, visibility, wind speed and direction, precipitation, cloud cover, and events for that day (i.e., rainy, foggy or sunny). The minimum, maximum, and mean of the first six attributes of the weather information were recorded in this data set. This data set was used to match the daily weather attributes with the first data set utilizing the two mutual attributes between them: date and ZIP code. The matched weather data was concatenated with the first data set.

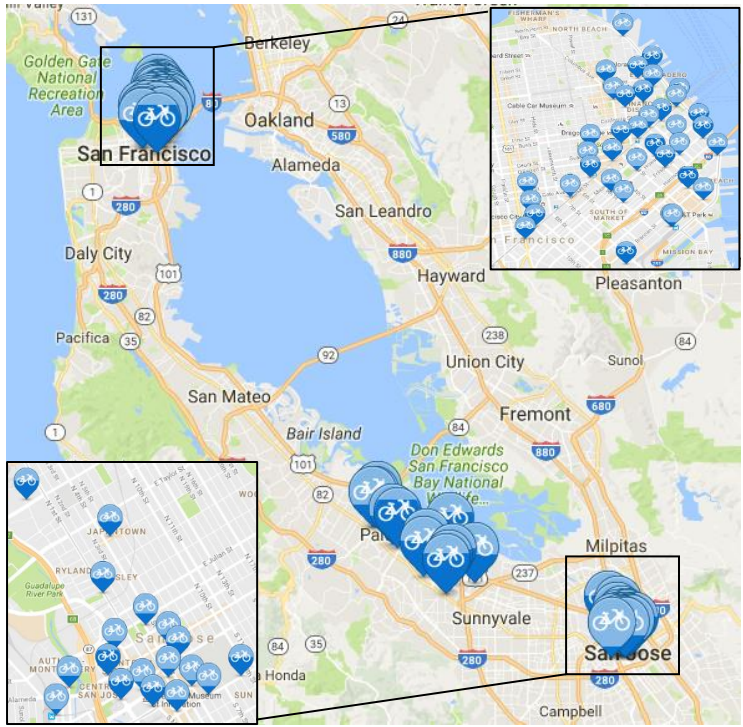

*Figure 2. Stations map [3]*

# 6 Data Analysis and Results

The following subsections present the methodology and the results of the data analysis. In implementing the count regression models—Poisson and Negative Binomial, RF, and BIC—MATLAB was used.

## 6.1 Problem Definition and Formulation

In quantifying the effect of various features on the system, we assumed that there is no interaction between the 70 stations and, thus station dummies were used to set up the model in this section for two reasons: (1) the main contribution of this section is to introduce an effective and fast, but also accurate and reasonable, approach to quantifying the effect of various features on bike counts at different stations. (2) One of the important contributions of this study is to investigate the possibility of pooling all of the variables in one model instead of developing a model for each of the 70 stations. This method could be reasonable and effective in cases of large networks with big data and various variables, and also, not needing a very high estimating accuracy at specific stations. This shall depend on the task and the level of accuracy that it is needed by the operator. In practice, operators at the strategic level would use the estimation of mean bike counts with no interactions between stations to cluster the stations and/or to determine if the number of docks is sufficient in some stations. Moreover, in some cases it would be used to predict the occupancy trends of the stations to improve the quality of the service and make it more reliable for the users [29].

As we assumed there was no interaction between the 70 stations, the $\log(\mu)$ of the bike count in each station might be represented as parallel hyperplanes. In order to construct one model containing all the stations instead of a model for each station, 69 indicator variables were coded as the 70 stations in the network, which implies that Station 1 is the reference in the model intercept. Similarly, 11 indicators were coded for the 12 months with January as a reference, six indicators for the seven days of the week were coded with Sunday as a reference, and two indicators for the events in the day were coded with sunny as a reference. All of these indicators were pooled in one model. If there was no significant difference between two of the parameters (say for example $\beta_1$ and $\beta_2$), this meant that the corresponding two parallel hyperplanes (Station 1 and Station 2) were very close to each other and the predicted $\log(\mu)$ of the bike count was the same for the two stations to an acceptable level of accuracy.

The first step in understanding the bike count's behavior was to regress all the available predictors to generate a full model. To that end, the PRM and NBRM were applied. The next step was using RF to rank the predictors in the full model based on the OOB error. Forward step-wise regression was then used to fit several models that were constructed as a result of the RF. Finally, BIC was used to select the best model, or, in other words, the best subset of predictors to construct this model.

However, this subset of predictors still had to be evaluated to determine whether they were reasonable. To accomplish this, all the parameters were examined and it was determined which were most acceptable. Different stations, month-of-the-year, day-of-the-week, and time-of-the-day were all determined to be reasonable parameters that might affect the model. From the weather information, mean temperature, mean humidity, mean visibility, mean wind speed, precipitation, and events were selected for further investigation. These parameters were selected based on subject-matter expertise, previous related studies (see for example [11, 20]), and to avoid multi-collinearity between two or more predictors. Once again, RF and forward step-wise regression were repeated and BIC was used to compare the built models. We chose the model with the best compromise between the minimum BIC value and the consideration of the effective parameters.

Two count models were used in this section: Poisson and negative binomial. To compare them, log-likelihood was estimated to determine goodness of fit. The likelihood of a set of parameter values is equal to the probability of the observed outcomes given those parameter values [30]. Table 1 shows the log-likelihood of Poisson and negative binomial for the full model. As negative binomial was able to accommodate the over-dispersion/under-dispersion in the data, its log-likelihood was higher than Poisson's. This meant that negative binomial was better than Poisson at describing the available bikes in the network. As a result, the NBRM was selected for use in all following steps in the analysis.

*Table 1. Log-likelihood of Poisson and negative binomial models*

|                | Poisson   | NB        |
|----------------|-----------|-----------|
| **Log-likelihood** | -5.95E+06 | -5.61E+06 |

## 6.2 Random Forest and Bayesian Information Criterion

Both RF and BIC were applied twice in this study. RF was applied on all the available predictors, constructing 111 different models. Basically, RF was implemented to sort the predictors in descending order of their relative "importance." MATLAB's manual describes this RF measurement as "*an array containing a measure of importance for each predictor variable (feature). For any variable, the measure is the increase in prediction error if the values of that variable are permuted across the out-of-bag observations. This measure is computed for every tree, then averaged over the entire ensemble and divided by the standard deviation over the entire ensemble*"[31]. Importance was utilized as a guide in forward step-wise regression using the NBRM, and computing the log-likelihood following each addition. BIC was then computed from the log-likelihood.

The BIC results of this first process are presented as the orange line shown in Figure 3. As the number of inserted predictors increased in the model, the BIC value decreased, indicating a better model. The BIC curve was used to select the most influential predictors resulting in the lowest BIC value. There was no specific rule for selecting those predictors, but rather it was a trade-off between the best and most simple model. The elbow in the curve, which corresponds to 45 predictors, was chosen to achieve the best compromise. The selected subset contained features of 31 stations, 7 months, 5 days, time-of-day, and one weather variable (wind direction degree). Based on subject-matter expertise and knowledge gained from related studies, it was determined that this subset was largely unacceptable. For example, temperature, not included in the subset, was found to be significant in previous studies of modeling bike counts in [11].

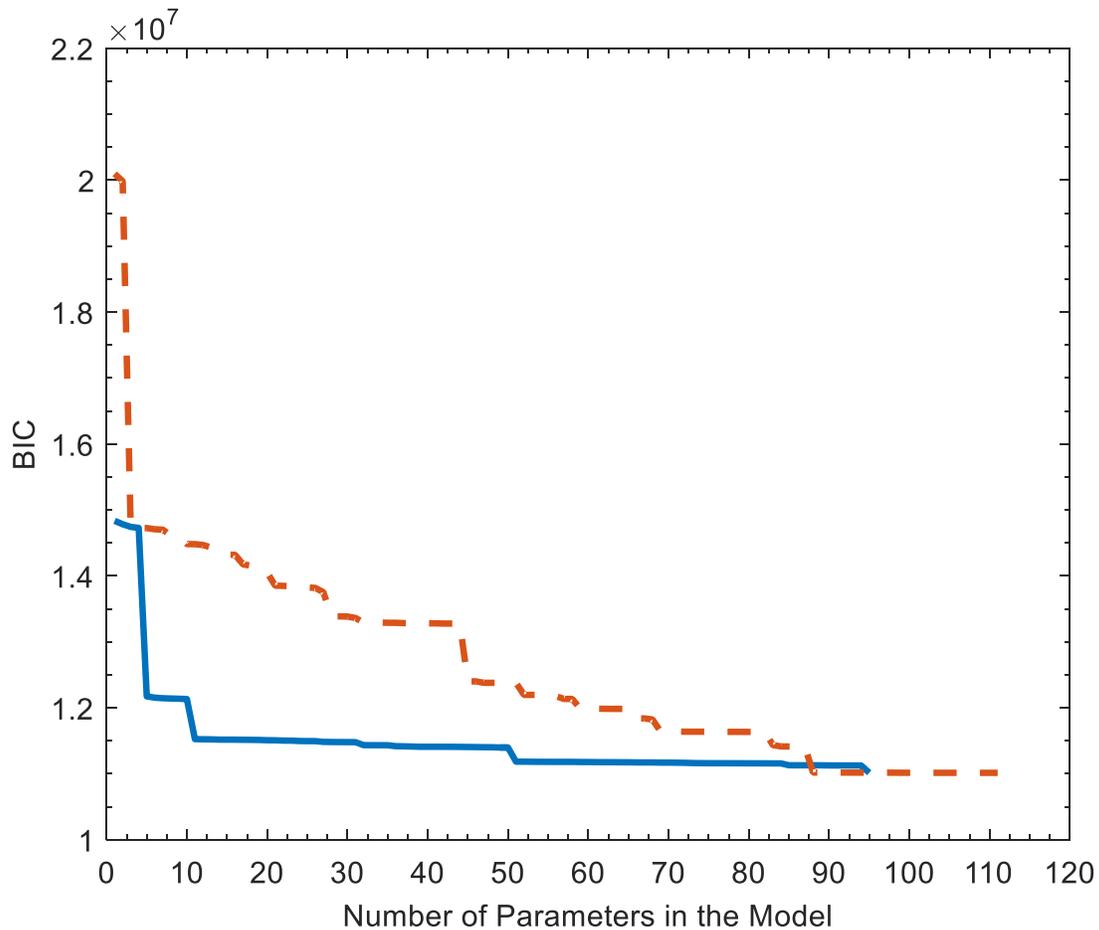

*Figure 3. BIC before (orange dashed line), and after (blue solid line) feature selection process*

This first conclusion led to a re-evaluation of the predictors by closely examining the weather information variables to determine any correlation among them. Again, based on expertise and related studies, mean temperature, mean humidity, mean visibility, mean wind speed, precipitation, and events were selected as predictors. RF and BIC were again applied after the predictor selection process. The importance of the predictors resulting from the RF is shown in Figure 4 (a) and the result of the BIC following forward step-wise regression is represented by the blue line in Figure 3. As Figure 3 illustrates, selecting these features improves BIC values remarkably. This is mainly because RF obtained a different order of predictors after neglecting any features that might correlate with other parameters. For example, maximum and minimum temperatures were correlated with the mean temperature. Maximum and minimum temperatures were neglected and the mean temperature remained.

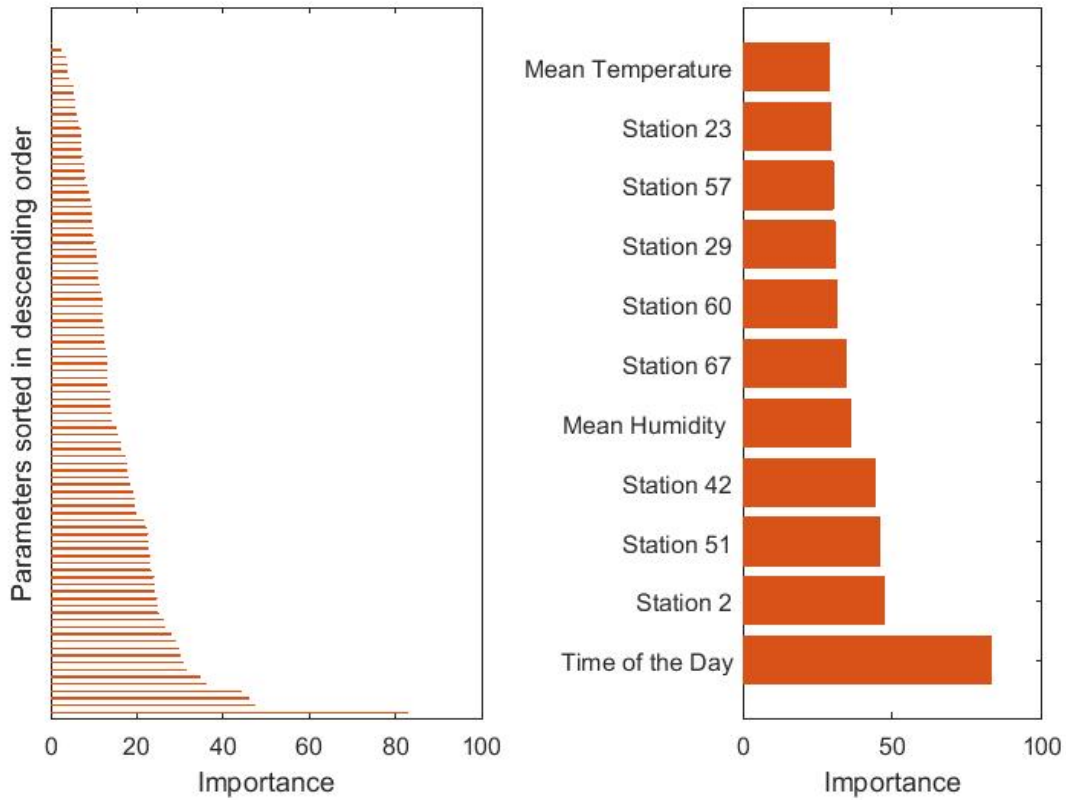

*Figure 4 Importance of the predictors (a) after feature selection (b) of the first proposed solution*

The BIC curve after feature selection revealed that two elbows could be selected as two proposed solutions that might achieve the best compromise: the first using 11 predictors, the second using 51 predictors. As the simplest explanation is preferable, the first solution was selected as the final model. Figure 4 (b) shows the importance of these 11 predictors, which are clearly reasonable. Temperature and humidity turned out to be important features and have significant effects in predicting bike availability in the Bay Area Bike Share network. Bay Area is one of the most humid areas in the United States, with an average humidity of nearly 74% [32]. Humidity has been proven to be a discomfort to people, particularly during physical activities like riding a bicycle.

Although we chose the first solution, it is worth noting that if we had selected the second solution, another two weather variables (visibility and wind speed), some days of the week, and some months would be included in the 51 most important predictors. All of these predictors are also reasonable and important in predicting bike availability in the Bay Area Bike Share network. The final model is formulated as follows:

$$\log(\mu) = \beta_0 + \beta_1 ToD + \beta_2\, S2 + \beta_3 S51 + \beta_4\, S42 + \beta_5 Hu + \beta_6\, S67 + \beta_7\, S60 \\ + \beta_8 S29 + \beta_9 S57 + \beta_{10}\, S23 + \beta_{11}\, T \qquad (6)$$

where:
$S$: Station,
$\beta_i$: Standardized coefficients,
$ToD$: Time − of − day,
$Hu$: Mean humidity (%),
$T$: Mean temperature ($F^o$)

This model is sufficient to estimate the mean number of bikes at each of the 70 stations producing relatively reasonable log-likelihood and BIC measures of -5.56E+06 and 1.12E+07, respectively. The log-likelihood for the reduced model is found to be higher (i.e. better) than the log-likelihood for the full model (see Table 1). When all the parameters in the full model were examined to determine which are most acceptable, we intended to exclude some parameters based on subject-matter expertise, previous related studies, and to avoid multicollinearity between the predictors, especially in the weather information. For example: mean, max, and min temperature were all regressed in the full model.

Multicollinearity in the full model is the cause of non-convergence or slow convergence of the maximum likelihood estimators, which means there is no longer a unique maximum point (i.e. peak) in the likelihood function; instead, there is a ridge [33, 34]. It appears that with collinear variables, the value of the parameter estimates fluctuates with no corresponding change in the log-likelihood. When we avoided the multicollinearity in the reduced model, the maximum likelihood estimator converged, and increased (i.e. improved) the corresponding log-likelihood.

Although the model was set up using station dummies, it does not imply that the model could only be used to estimate the mean bike counts for the entire network. Rather, this model can be used to estimate the mean bike counts at each of the 70 stations. If one is interested in estimating the mean bike counts at Station 60, for example, then the model will be:

$$\log(\mu) = \beta_0 + \beta_1 ToD + \beta_5 Hu + \beta_7\, S60 + \beta_{11}\, T \qquad (7)$$

and all other station covariates in the model equal zero. However, if one would like to estimate the mean bike counts at Station 50 that is not included in the reduced model, then the model will be:

$$\log(\mu) = \beta_0 + \beta_1 ToD + \beta_5 Hu + \beta_{11}\, T \qquad (8)$$

This implies two inferences, as follows: (1) There is no significant effect in including the station parameter in the model given that the mean bike counts at Station 50 is determined by three variables, namely: the time-of-the-day, the humidity level, and the ambient temperature. In other words, there is no significant difference between Station 50 and the reference station parametrized in the interception (i.e. Station 1) (2) There is no considerable difference between estimating the mean bike counts of Station 50 and, for example, Station 40 (also not included in the model), for the same time-of-the-day, humidity level, and ambient temperature. This is because the corresponding two parallel hyperplanes for Station 50 and Station 40 are very close to each other and the estimated $log(\mu)$ of the mean of bike counts is the same for the two stations to an acceptable level of accuracy.

Table 2 shows the estimated parameter values for the NB Model of mean bike counts in the studied network. It shows also that all the parameters are significant since the p-values are approximately equal to zero.

*Table 2. Estimated parameter values for the NB model for bike availability in the network*

|  | **Estimate** | **P-value** |
|---|---|---|
| **Intercept** | 2.226865 | $< 0.0001$ |
| **Time-of-the-day** | -0.00050 | $< 0.0001$ |
| **Station 2** | 0.467929 | $< 0.0001$ |
| **Station 51** | 0.411846 | $< 0.0001$ |
| **Station 42** | 0.290969 | $< 0.0001$ |
| **Humidity** | 0.000516 | $< 0.0001$ |
| **Station 67** | 0.428846 | $< 0.0001$ |
| **Station 60** | 0.186177 | $< 0.0001$ |
| **Station 29** | 2.56E-01 | $< 0.0001$ |
| **Station 57** | 0.217112 | $< 0.0001$ |
| **Station 23** | 0.290833 | $< 0.0001$ |
| **Temperature** | -0.0013 | $< 0.0001$ |

## 6.3 Bike Count Modeling for Each Station

In this section, we use the proposed approach to modeling bike counts at each bike-sharing station using the NBRM as it appeared to outperform the PRM. Since the number of available bikes at a station, which has a finite number of docks, fluctuates, a repositioning (or redistribution) operation must be performed periodically. Coordinating such a large operation is complicated, time-consuming, polluting, and expensive [1]. Modeling the bike count at each station considering various features is one of the key tasks to making this operation more efficient. This task is the full prediction problem that would help planners make decisions such as determining the stations that need rebalancing over the entire day, and considering relocation of underused stations (or building new ones) to serve busy areas in the network. NBRM was applied to create predictive models to predict the bike counts at each of the 70 stations of the Bay Area BSS network. For each model, we used the proposed method to quantify the effect of different variables and then selected the most accurate model while maintaining model simplicity. The RF was used to rank the effect of the different parameters in the model. This rank was used as a systematic guide in the forward step-wise regression. The subset of variables includes 25 features for each station, including: month-of-the-year, day-of-the-week, time-of-the-day, and some weather variables. The weather information contains mean temperature, mean humidity level, mean visibility level, mean wind speed, precipitation intensity, and events of fog and snow.

In [35, 36], we studied the effect of using bike count memory data at station $i$ as a prediction variable by ranging the memory (of 15 minutes) from $t-1$ to $t-7$, in which $t$ is the model without including any memory data. Results showed that memory data beyond $t-1$ had a relatively small effect on bike count prediction. It seems that they do not add further explanation for the response's variability. As a result, in addition to the abovementioned subset of variables, we also added the number of available bikes at station $i$ at time $t-1$ to estimate the bike counts at each station $i$ in the network at time $t$.

Figure 5 shows the mean prediction error ($MPE$) for a randomly selected test sample that was not used in the estimation process for all the stations using the proposed method. As Figure 5 shows, the first 32 stations and the last two stations have relatively lower $MPE$ than the other 36 stations. In this BSS, there are 70 bike stations that connect users in four main areas: downtown San Francisco, Palo Alto, Mountain View, and downtown San Jose. Stations that have relatively higher $MPE$ are based in downtown San Francisco, which has a population density approximately 10 times higher than the population of the other three areas [37]. Moreover, we hypothesize that people tend to use public transportation, including BSS, in San Francisco more than the other three areas. The annual report of the TomTom Traffic Index of 2017 [38] indicates that drivers in San Francisco incur an average of 39% extra travel time while stuck in traffic anytime of the day, which is 7% more than what San Jose's drivers experience. This suggests that demographic and built environment variables are critical factors in predicting bike counts.

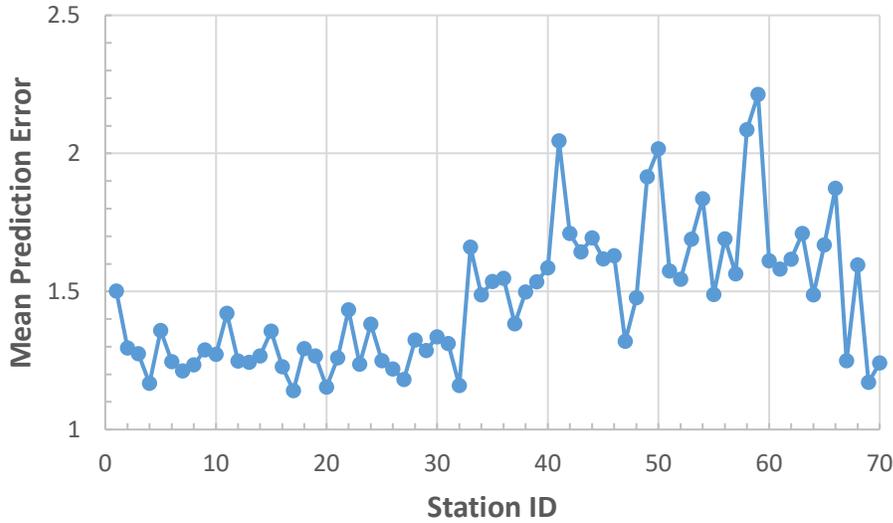

Figure 5. MPE at each station using the proposed method

Although we will present the results of modeling bike counts only at Station 3 for illustration purposes, we ran the proposed method for all the stations and the results were consistent with the presented results in terms of the variables chosen by the proposed method. The BIC curve for Station 3 in Figure 6 (a) revealed different elbows that could be selected as a proposed model. To achieve a good compromise between the BIC and the simplicity of the model, three parameters were selected to be part of the final model. The prediction results for the final model for station 3, which is shown in Figure 6 (b), is formulated as follows:

$$\log(\mu_{S3}) = \beta_0 + \beta_1\, Y_{t-1} + \beta_2\, ToD + \beta_3\, Hu \qquad (9)$$

where:
$S$: $Station$,
$\beta_0 = 1.24, \beta_1 = 0.1025, \beta_2 = -0.02, and\ \beta_3 = -0.005$,
$Y_{t-1}$: $bike\ count\ memory\ data\ at\ t-1\ (15\ minutes\ ago)$,
$ToD$: $Time-of-day$,
$Hu$: $Mean\ humidity\ (\%)$,

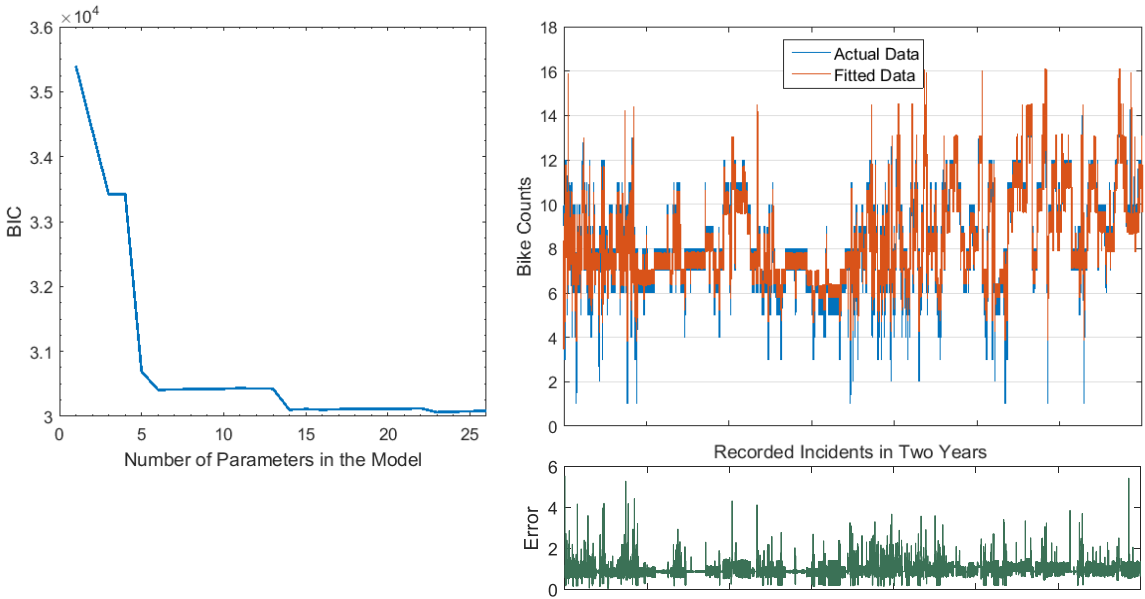

*Figure 6. (a) BIC curve, and (b) fitted model for Station 3*

# 7 Conclusions and Recommendations for Future Work

In this paper, we described the development of a bike availability model for the San Francisco Bay Area Bike Share program. Since the bike count estimation and prediction are still not well studied, this paper introduced an effective and fast, but also accurate and reasonable, approach to quantifying the effect of various features on bike counts at different stations. The results revealed that the bike counts changed with the month-of-the-year, day-of-the-week, time-of-the-day, and some weather variables.

NBRM and PRM were performed on the bike count data. The NBRM was ultimately chosen, as it was found to best fit the count data. However, the significance measure in NBRM (i.e., p-value) resulting from directly regressing all variables is not always an adequate measure and also depends on the order of parameters being regressed in the model, especially when there is a large number of features and if there is a possible correlation between these features. As a result, this study adopted a new method consisting of feature selection using RF. RF was run on the predictors guiding a forward step-wise regression and using the BIC to compare models. This method turned out to be an effective and reasonable approach to identify critical predictors of bike counts in the system and at each station.

The final results reveal interesting new insights. Firstly, this is the first study to use the mean humidity level as a predictor of bike counts. Results of this study demonstrate that humidity is a significant predictor in the Bay Area Bike Share program. Further, although the precipitation level has been shown to be significant in many previous studies, the results of this study demonstrate that precipitation is not a significant predictor in the Bay Area. Over the entire year, the most common forms of precipitation in the Bay Area were light rain, moderate rain, and drizzle, none of which appeared to have a major effect on Bay Area Bike Share use. The contrast between this finding and that of previous studies indicates that particular weather information may have different significance depending on the studied geographic area.

Secondly, in investigating the effect of variables in the BSS, eight indicator variables corresponding to eight stations and one variable serving as a reference in the intercept were selected as final estimators in the model. This implies that the mean bike count data for the remaining 61 indicator variables corresponding to 61 stations are not significantly different from the mean bike count data for the reference station. The variability in bike counts of these 61 stations would not be influential if the data were employed as estimators in the regression. Nonetheless, the eight stations were different from the reference station to an extent that might largely affect the estimation if they are not considered. This is because of these station locations. For example, one station is near the main train station in Palo Alto, which is the second busiest station in the Caltrain system; another is near Yerba Buena Center for the Arts in San Francisco; one is at Union Square, which is a busy public square in the center of San Jose; and one is at the San Antonio Caltrain station in Mountain View.

Finally, the number of available bikes at station $i$ at time $t-1$ and the time-of-the-day were found to be of the most important predictors in modeling the bike counts at each station. This means that the bike count fluctuates over the course of the day (i.e., during peak and off-peak periods). The constructed models for each station could also be used to improve the redistribution of bicycles, which is important for rebalancing the network over a period of time.

Although the adopted approach needs to be further validated by applying it to other bike count data in different geographic areas, results demonstrate that it is promising in quantifying the effect of various features in cases of large networks with big data. It is also important in the future to investigate other variables such as bikes coming from other stations and the relative location of each station.

# 8   Acknowledgements

This effort was funded by the Urban Mobility and Equitable Center and the National Science Foundation UrbComp project. The authors also acknowledge the assistance of Ahmed Ghanem and Mohammed Almannaa in data reduction and cleaning.